\documentclass[twocolumn,aps,prl,final,superscriptaddress]{revtex4-1}
\usepackage[latin9]{inputenc}
\usepackage[active]{srcltx}
\usepackage{color}
\usepackage{multirow}
\usepackage{amsmath}
\usepackage{graphicx}
\usepackage[unicode=true,pdfusetitle,
 bookmarks=true,bookmarksnumbered=true,bookmarksopen=true,bookmarksopenlevel=1,
 breaklinks=true,pdfborder={0 0 1},backref=false,colorlinks=true]
 {hyperref}
\hypersetup{
 linkcolor=blue,urlcolor=blue,citecolor=blue,pdfstartview={FitH},hyperfootnotes=false}

\makeatletter

\newcommand*\LyXThinSpace{\,\hspace{0pt}}
\providecommand{\tabularnewline}{\\}

\usepackage{dcolumn}
\usepackage{bm}

\usepackage{times}

\makeatother

\begin{document}
\title{Giant electro-optic and elasto-optic effects in ferroelectric NbOI$_{2}$}
\author{Zhenlong Zhang}
\affiliation{Ministry of Education Key Laboratory for Nonequilibrium Synthesis
and Modulation of Condensed Matter, Shaanxi Province Key Laboratory
of Advanced Functional Materials and Mesoscopic Physics, School of
Physics, Xi'an Jiaotong University, Xi'an 710049, China}
\affiliation{State Key Laboratory of Surface Physics and Department of Physics,
Fudan University, Shanghai 200433, China}
\author{Xuehan Di}
\affiliation{Ministry of Education Key Laboratory for Nonequilibrium Synthesis
and Modulation of Condensed Matter, Shaanxi Province Key Laboratory
of Advanced Functional Materials and Mesoscopic Physics, School of
Physics, Xi'an Jiaotong University, Xi'an 710049, China}
\author{Charles Paillard}
\affiliation{Smart Ferroic Materials Center, Physics Department and Institute for
Nanoscience and Engineering, University of Arkansas, Fayetteville,
Arkansas 72701, USA}
\affiliation{Université Paris-Saclay, CentraleSupélec, CNRS, Laboratoire SPMS,
91190, Gif-sur-Yvette, France}
\author{Laurent Bellaiche}
\email{laurent@uark.edu}

\affiliation{Smart Ferroic Materials Center, Physics Department and Institute for
Nanoscience and Engineering, University of Arkansas, Fayetteville,
Arkansas 72701, USA}
\affiliation{Department of Materials Science and Engineering, Tel Aviv University,
Ramat Aviv, Tel Aviv 6997801, Israel}
\author{Zhijun Jiang}
\email{zjjiang@xjtu.edu.cn}

\affiliation{Ministry of Education Key Laboratory for Nonequilibrium Synthesis
and Modulation of Condensed Matter, Shaanxi Province Key Laboratory
of Advanced Functional Materials and Mesoscopic Physics, School of
Physics, Xi'an Jiaotong University, Xi'an 710049, China}
\affiliation{State Key Laboratory of Surface Physics and Department of Physics,
Fudan University, Shanghai 200433, China}
\begin{abstract}
First-principles calculations are performed to investigate the electro-optic
(EO) and elasto-optic effects of the three-dimensional (bulk) and
two-dimensional (monolayer) ferroelectric NbOI$_{2}$. Remarkably
large linear EO and elasto-optic coefficients are discovered in both
systems, when under stress-free conditions. We further found that
the EO responses of bulk and monolayer NbOI$_{2}$ can be further
enhanced with epitaxial strain, because of a strain-driven ferroelectric-to-paraelectric
transition that originates from the softening of some polar optical
modes. Our findings thus point out that NbOI$_{2}$, as well as other
niobium oxide dihalides are highly promising for paving the way for
potentially efficient nonlinear optical device applications. 
\end{abstract}
\maketitle
The linear electro-optic (EO) effect (or Pockels effect) is defined
as the variation of the refractive index of a material affected by
an external electrical field \cite{DiDomenico1969,Wemple1969,Weber2002}.
It has the potential to be used for power-efficient and high-speed
optical devices \cite{Turner1966,Lines1997,Wemple1972,Saleh1991,Yariv2007,Boyd2008,Bass2009},
such as EO modulators \cite{Xu2005}, bistable switches \cite{Lorente2017},
and optical resonators \cite{Guarino2007}. However, there are very
few materials possessing large linear EO effect, which limits its
application. The standard material of LiNbO$_{3}$ is currently the
best choice for optical modulators in the telecommunications industry
because of its large linear EO coefficient ($\sim$ 30 pm/V) \cite{Weber2002,Turner1966}.
Finding other materials with even larger EO coefficient is therefore
of high importance for practical applications but also for scientific
interest by revealing the microscopic reason behind such hypothetical
enhancement.

Recently, first-principles calculations predicted very large ferroelectric
and piezoelectric effects in the NbOX$_{2}$ systems \cite{Jia2019,Ye2021,Wu2022}.
These layered niobium oxide dihalides NbOX$_{2}$ (X=Cl, Br, I) systems
were also discovered to exhibit large second-harmonic generation (SHG)
likely due to the interplay between anisotropic polarization and excitonic
resonance \cite{Fang2021,Abdelwahab2022,Fu2023,Ye2023,Guo2023,Wang2024,Yan2024,Chen2024}.
Note that the SHG intensity of NbOX$_{2}$ family is proportional
to the ferroelectric spontaneous polarization, which is highly promising
for exploring large second-order optical nonlinearities. However,
to the best of our knowledge, electro-optic effects, but also elasto-optic
conversion that relates a change of strain with a variation in the
refractive index, remain unknown in these materials.

The aim of this Letter is to investigate the linear electro-optic
effect and elasto-optic effects in stress-free ferroelectric NbOI$_{2}$
bulk and monolayer, as well as the effect of epitaxial strain on such
coupling properties in these systems. As we will see, both NbOI$_{2}$
bulk and monolayer possess really large EO and elasto-optic coefficients
for stress-free conditions, with these responses becoming even giant
for some epitaxial strains. The reason behind such enhancement of
these non-linear effects is further revealed.

Here, we choose the ground state structures of ferroelectric NbOI$_{2}$
bulk and monolayer, which have the $C2$ space group ($2$ point group)
and $Pmm2$ space group ($mm2$ point group), respectively. First-principles
calculations are performed on the ferroelectric bulk and monolayer
structures based on the density functional theory (DFT) with the generalized
gradient approximation of the Perdew-Burke-Ernzerhof (PBE) exchange-correlation
functional form, using the ABINIT package \cite{Gonze2002} with the
norm-conserving pseudopotentials \cite{Hamann2013,Setten2018}. We
use a $\Gamma$-centered 12$\times$6$\times$1 $k$-point mesh to
sample the Brillouin zone of bulk and monolayer NbOI$_{2}$ and a
plane-wave cutoff of 50 hartrees. For the ferroelectric monolayer
NbOI$_{2}$, a vacuum space of more than 20 Å is used to avoid the
periodic image interactions. Furthermore, the effects of epitaxial
biaxial strain on the structural properties of bulk and monolayer
NbOI$_{2}$ are calculated as well. The considered strains are ranging
between $-$3\% and $+$3\%. For each considered strain, the in-plane
lattice vectors are kept fixed, while the out-of-plane lattice vector
is allowed to relax for bulk NbOI$_{2}$ but kept fixed for the supercell
modeling the monolayer and vacuum. The atomic positions are fully
relaxed for both the bulk and monolayer NbOI$_{2}$ until all the
force values acting on the atoms are less than 1$\times$10$^{-6}$
hartree/bohr. In order to mimic the van der Waals (vdW) interactions
in NbOI$_{2}$ bulk, the DFT-D3 method with Becke-Johnson damping
\cite{Grimme2011} was included in the calculations. The strain-induced
structures are then used to calculate the EO tensor but within the
local density approximation (LDA) because such calculation is only
implemented within LDA. This method can be rather accurate, since
it predicts linear and nonlinear EO coefficients for ferroelectric
oxides \cite{Veithen2004,Veithen2005,Jiang2020} that agree rather
well with experimental results \cite{Weber2002,Chen2014}. Technically,
the linear EO tensor ${r}_{ijk}$ is expressed as: 
\begin{equation}
\Delta(\varepsilon^{-1})_{ij}=\sum_{k=1}^{3}{r}_{ijk}{\cal E}_{k},\label{eq:EO tensor}
\end{equation}
where $(\varepsilon^{-1})_{ij}$ is the inverse of the electronic
dielectric tensor and ${\cal E}_{k}$ is the applied electric field
in the Cartesian direction $k$.

The \textit{clamped} (strain-free) EO tensor can be written as \cite{Veithen2004,Veithen2005}:
\begin{equation}
{r}_{ijk}^{\mathrm{\eta}}={r}_{ijk}^{\mathrm{\textrm{el}}}+{r}_{ijk}^{\mathrm{\textrm{ion}}}=\frac{-8\pi}{n_{i}^{2}n_{j}^{2}}\chi_{ijk}^{(2)}-\frac{4\pi}{n_{i}^{2}n_{j}^{2}\sqrt{\Omega_{0}}}\sum_{m}\frac{\alpha_{ij}^{m}p_{k}^{m}}{\omega_{m}^{2}},\label{eq:EO_tensor_supercell}
\end{equation}
where $r_{ijk}^{\mathrm{el}}$ is the bare electronic contribution,
$r_{ijk}^{\mathrm{ion}}$ represents the ionic contribution, $n_{i}$
and $n_{j}$ are the principal refractive indices, $\chi_{ijk}^{(2)}$
is the nonlinear optical dielectric susceptibility, $\Omega_{0}$
is the unit cell volume, $\alpha_{ij}^{m}$ denotes the Raman susceptibility
of mode $m$, $p_{k}^{m}$ is the polarity, and $\omega_{m}$ is the
phonon mode frequency. Note that the clamped EO tensor and phonon
frequencies are directly obtained from density functional perturbation
theory (DFPT) calculations \cite{Veithen2005}.

The \textit{unclamped} (stress-free that adds a piezoelectric contribution)
EO coefficients can be expressed as

\begin{equation}
{r}_{ijk}^{\mathrm{\sigma}}=r_{ijk}^{\eta}+\sum_{\alpha,\beta}p_{ij\alpha\beta}d_{k\alpha\beta},\label{eq:unclamped EO coefficient}
\end{equation}
where $p_{ij\alpha\beta}$ is the elasto-optic coefficients and $d_{k\alpha\beta}$
represent the piezoelectric strain coefficients. Note that Eqs.~(\ref{eq:EO_tensor_supercell})
and (\ref{eq:unclamped EO coefficient}) can be used to calculate
the linear clamped and unclamped EO tensor from first-principles \cite{Charles2019,Jiang2019,Paoletta2021,Prosandeev2024}.
Measurements with time response method (TRM) \cite{Abarkan2014} and
modulation depth method (MDM) \cite{Abarkan2017} can be used to determine
the clamped and unclamped EO coefficients.

Moreover, the elasto-optic tensor $p_{ij\alpha\beta}$ is given by
the expression \cite{Zgonik1994}

\begin{equation}
\Delta\left(\frac{1}{n^{2}}\right)_{ij}=\Delta\left(\varepsilon^{-1}\right)_{ij}=\sum_{\alpha,\beta}p_{ij\alpha\beta}x_{\alpha\beta},\label{eq:elasto-optic coefficient}
\end{equation}
where $n$ is the refractive index (which is simply equal to the square
root of the electronic dielectric tensor) and $x_{\alpha\beta}$ denotes
the strain tensor.

We first check the structural parameters of NbOI$_{2}$ bulk and monolayer.
The lattice parameters of bulk NbOI$_{2}$ are found to be $a$ $=$
3.89 Å, $b$ $=$ 7.51 Å, $c$ $=$ 15.15 Å and $\alpha$ $=$ 105.4${^{\circ}}$,
which are in good agreement with the experimental values of $a$ $=$
3.92 Å, $b$ $=$ 7.52 Å, $c$ $=$ 15.18 Å, and $\alpha$ $=$ 105.5${^{\circ}}$
\cite{Rijnsdorp1978}. For ferroelectric NbOI$_{2}$ monolayer, the
relaxed lattice constants are $a$ $=$ 3.94 Å, $b$ $=$ 7.58 Å,
which are precisely those found in Refs.~\cite{Ye2021,Ye2023}.

\begin{figure}
\includegraphics[width=8.5cm]{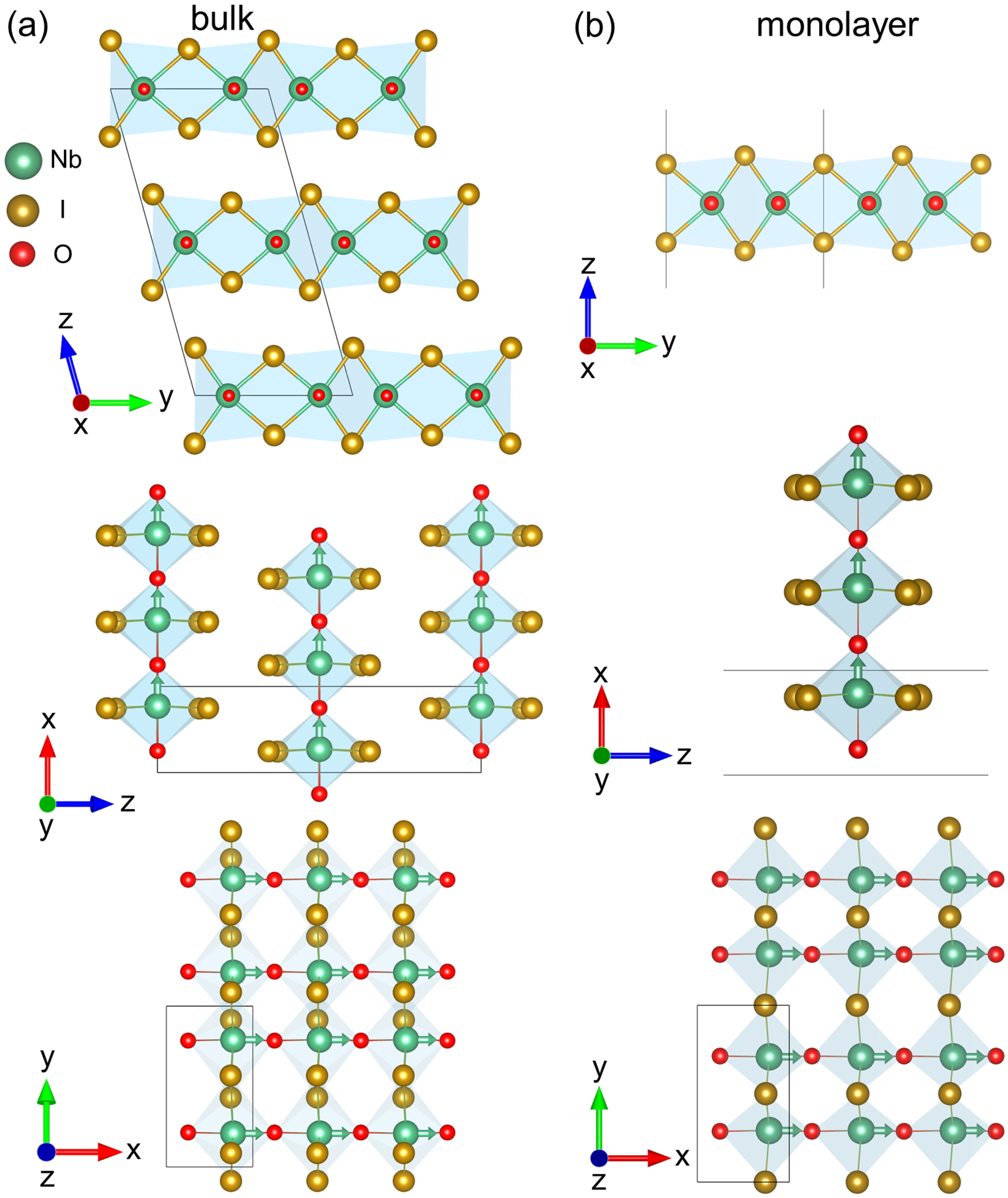}

\caption{Crystal structures of ferroelectric NbOI$_{2}$ bulk and monolayer.
Panels (a) and (b) show the side and top views of NbOI$_{2}$ bulk
and monolayer, respectively. Black lines represent the simulated unit
cell. The arrows centred on the Nb ions represent the spontaneous
polarization direction. \label{fig:Crystal structures }}
\end{figure}

Figure~\ref{fig:Crystal structures } displays the crystal structures
of the NbOI$_{2}$ bulk and monolayer, respectively. A spontaneous
polarization occurs along the $x$-axis direction for both ferroelectric
NbOI$_{2}$ bulk and monolayer. The EO tensor of NbOI$_{2}$ bulk
($2$ point group) has eight independent elements in the Voigt notation
\cite{Nye1985}: ${r}_{11}$, ${r}_{21}$, ${r}_{31}$, ${r}_{41}$,
${r}_{52}$, ${r}_{53}$, ${r}_{62}$ and ${r}_{63}$, while the EO
tensor of the NbOI$_{2}$ monolayer ($mm2$ point group) has five
independent elements: ${r}_{11}$, ${r}_{21}$, ${r}_{31}$, ${r}_{53}$
and ${r}_{62}$. For the NbOI$_{2}$ monolayer, it is important to
recall that the calculated EO tensor predicted by the DFT simulations
on the supercell has to be rescaled as \cite{Jiang2022,Jiang2024}:
\begin{equation}
r_{11}^{\textrm{2\ensuremath{\mathrm{D}}}}=\frac{c}{t}\frac{\left(\varepsilon_{\textrm{11}}^{\textrm{SC}}\right)^{2}}{\left(\varepsilon_{\textrm{11}}^{\textrm{2\ensuremath{\mathrm{D}}}}\right)^{2}}r_{11}^{\textrm{SC}},\label{eq:r11_2D}
\end{equation}

\begin{equation}
r_{21}^{\textrm{2\ensuremath{\mathrm{D}}}}=\frac{c}{t}\frac{\left(\varepsilon_{\textrm{22}}^{\textrm{SC}}\right)^{2}}{\left(\varepsilon_{\textrm{22}}^{\textrm{2\ensuremath{\mathrm{D}}}}\right)^{2}}r_{21}^{\textrm{SC}},\label{eq:r21_2D}
\end{equation}

\begin{equation}
r_{\textrm{31}}^{\textrm{2\ensuremath{\mathrm{D}}}}=\frac{c}{t}r_{\textrm{31}}^{\textrm{SC}},\label{eq:r31_2D}
\end{equation}

\begin{equation}
r_{53}^{\textrm{2\ensuremath{\mathrm{D}}}}=\frac{\varepsilon_{\textrm{11}}^{\textrm{SC}}}{\varepsilon_{\textrm{11}}^{\textrm{2D}}}r_{53}^{\textrm{SC}},\label{eq:19}
\end{equation}

\begin{equation}
r_{62}^{\textrm{2\ensuremath{\mathrm{D}}}}=\frac{c}{t}\frac{\varepsilon_{\textrm{11}}^{\textrm{SC}}\varepsilon_{\textrm{22}}^{\textrm{SC}}}{\varepsilon_{\textrm{11}}^{\textrm{2D}}\varepsilon_{\textrm{22}}^{\textrm{2D}}}r_{62}^{\textrm{SC}},\label{eq:23}
\end{equation}
where $r_{11}^{\textrm{2\ensuremath{\mathrm{D}}}}$, $r_{21}^{\textrm{2\ensuremath{\mathrm{D}}}}$,
$r_{31}^{\textrm{2\ensuremath{\mathrm{D}}}}$, $r_{53}^{\textrm{2\ensuremath{\mathrm{D}}}}$
and $r_{62}^{\textrm{2\ensuremath{\mathrm{D}}}}$ are the rescaled
2D EO coefficients; $r_{11}^{\textrm{SC}}$, $r_{21}^{\textrm{SC}}$,
$r_{31}^{\textrm{SC}}$, $r_{53}^{\textrm{SC}}$ and $r_{62}^{\textrm{SC}}$
are the EO tensor of the supercell; $\varepsilon_{\textrm{11}}^{\textrm{SC}}$,
$\varepsilon_{\textrm{22}}^{\textrm{SC}}$ and $\varepsilon_{\textrm{33}}^{\textrm{SC}}$
are the diagonal elements of the dielectric tensor in the supercell;
$\varepsilon_{\textrm{11}}^{\textrm{2D}}$, $\varepsilon_{\textrm{22}}^{\textrm{2D}}$
and $\varepsilon_{\textrm{33}}^{\textrm{2D}}$ are the renormalized
2D dielectric constants; $c$ is the lattice constant of the supercell
in monolayer NbOI$_{2}$ while $t$ is the effective thickness of
the 2D material. The choice of $t$ is based on the van der Waals
bond length \cite{Jiang2024,Laturia2018}. Practically, $t$ is taken
to be equal to 7.3 $\textrm{Å}$ here for the NbOI$_{2}$ monolayer,
which is consistent with the experimental report of Ref.~\cite{Abdelwahab2022}.
The rescaling of the dielectric constants is given by \cite{Laturia2018}:
$\varepsilon_{\textrm{11}}^{\textrm{2\ensuremath{\mathrm{D}}}}=1+\frac{c}{t}\left(\varepsilon_{\textrm{11}}^{\textrm{SC}}-1\right)$,
$\varepsilon_{\textrm{22}}^{\textrm{2\ensuremath{\mathrm{D}}}}=1+\frac{c}{t}\left(\varepsilon_{\textrm{22}}^{\textrm{SC}}-1\right)$,
and $\varepsilon_{\textrm{33}}^{\textrm{2\ensuremath{\mathrm{D}}}}=\left[1+\frac{c}{t}\left(\left(\varepsilon_{\textrm{33}}^{\textrm{SC}}\right)^{-1}-1\right)\right]^{-1}$.
Note also that the elasto-optic tensor of the monolayer needs to be
rescaled in the same way as the EO tensor in 2D systems \cite{Jiang2024}
while the piezoelectric strain coefficients do not need to because
they are independent of the thickness of the vacuum layers \cite{Wu2022}.

\begin{table}
\caption{\label{tab:Table-I} Clamped electro-optic coefficients in ferroelectric
NbOI$_{2}$ bulk and monolayer. \label{tab:clamped EO tensor}}

\begin{ruledtabular}
\begin{tabular}{cc}
bulk EO tensor (pm/V)  & monolayer EO tensor (pm/V)\tabularnewline
\hline 
\multirow{6}{*}{$\left[\begin{array}{ccc}
58.63 & 0 & 0\\
16.17 & 0 & 0\\
7.69 & 0 & 0\\
0.11 & 0 & 0\\
0 & -0.18 & 0.004\\
0 & 1.24 & -0.09
\end{array}\right]$} & \multirow{6}{*}{$\left[\begin{array}{ccc}
35.49 & 0 & 0\\
8.27 & 0 & 0\\
3.98 & 0 & 0\\
0 & 0 & 0\\
0 & 0 & 0.01\\
0 & 2.36 & 0
\end{array}\right]$}\tabularnewline
 & \tabularnewline
 & \tabularnewline
 & \tabularnewline
 & \tabularnewline
 & \tabularnewline
\end{tabular}
\end{ruledtabular}

\end{table}

\begin{table}
\caption{\label{tab:table-II} Unclamped electro-optic and elasto-optic coefficients
in ferroelectric NbOI$_{2}$ bulk and monolayer. \label{tab:unclamped EO tensor}}

\begin{ruledtabular}
\begin{tabular}{ccccccc}
 & \multicolumn{3}{c}{Unclamped EO tensor (pm/V)} & \multicolumn{3}{c}{Elasto-optic tensor}\tabularnewline
\hline 
NbOI$_{2}$  & $r_{11}^{\sigma}$  & $r_{21}^{\sigma}$  & $r_{31}^{\sigma}$  & $p_{11}$  & $p_{21}$  & $p_{31}$\tabularnewline
bulk  & 289.76  & 7.05  & $-$12.66  & 1.65  & 0.61  & 0.39\tabularnewline
monolayer  & 133.63  & 4.96  & 3.48  & 1.58  & 0.46  & 0.39\tabularnewline
\end{tabular}
\end{ruledtabular}

\end{table}

Table \ref{tab:clamped EO tensor} shows the clamped electro-optic
tensor in ferroelectric NbOI$_{2}$ bulk and monolayer. Regarding
NbOI$_{2}$ bulk, the predicted clamped EO coefficients are: $r_{11}^{\textrm{\ensuremath{\mathrm{\eta}}}}$
$=$ 58.63 pm/V, $r_{21}^{\textrm{\ensuremath{\mathrm{\eta}}}}$ $=$
16.17 pm/V, $r_{31}^{\textrm{\ensuremath{\mathrm{\eta}}}}$ $=$ 7.69
pm/V, $r_{41}^{\textrm{\ensuremath{\mathrm{\eta}}}}$ $=$ 0.11 pm/V,
$r_{52}^{\textrm{\ensuremath{\mathrm{\eta}}}}$ $=$ $-$0.18 pm/V,
$r_{53}^{\textrm{\ensuremath{\mathrm{\eta}}}}$ $=$ 0.004 pm/V, $r_{62}^{\textrm{\ensuremath{\mathrm{\eta}}}}$
$=$ 1.24 pm/V, and $r_{63}^{\textrm{\ensuremath{\mathrm{\eta}}}}$
$=$ $-$0.09 pm/V. The largest clamped EO coefficient in NbOI$_{2}$
bulk is therefore about twice larger than the currently most used
EO material LiNbO$_{3}$ which has a $r_{33}^{\textrm{\ensuremath{\mathrm{\eta}}}}$
$=$ 30.8 pm/V \cite{Weber2002,Veithen2004}. For NbOI$_{2}$ monolayer,
the clamped EO coefficients of $r_{11}^{\textrm{\ensuremath{\eta},2\ensuremath{\mathrm{D}}}}$
$=$ 35.49 pm/V, $r_{21}^{\textrm{\ensuremath{\eta},2\ensuremath{\mathrm{D}}}}$
$=$ 8.27 pm/V, and $r_{31}^{\textrm{\ensuremath{\eta},2\ensuremath{\mathrm{D}}}}$
$=$ 3.98 pm/V, are smaller in magnitude than those of bulk. In contrast,
the clamped EO coefficients $r_{53}^{\textrm{\ensuremath{\mathrm{\eta,2\ensuremath{\mathrm{D}}}}}}$
and $r_{62}^{\textrm{\ensuremath{\mathrm{\eta,2\ensuremath{\mathrm{D}}}}}}$
of monolayer are larger in magnitude than those of the bulk case but
both have small values. Note that the predicted EO coefficients $r_{11}^{\textrm{\ensuremath{\eta},2\ensuremath{\mathrm{D}}}}$
in NbOI$_{2}$ monolayer is also larger than the experimental results
in LiNbO$_{3}$ \cite{Weber2002,Turner1966} and 2D ferroelectric
CuInP$_{2}$S$_{6}$ (with an effective linear EO coefficient to be
20.28 pm/V) \cite{Liu2024}, which emphasizes the high potential of
NbOI$_{2}$ systems for the design of efficient electro-optic devices.

Table \ref{tab:unclamped EO tensor} displays some selected unclamped
EO coefficients and elasto-optic coefficients of NbOI$_{2}$ bulk
and monolayer, respectively. For NbOI$_{2}$ bulk, the largest unclamped
EO coefficient of $r_{11}^{\sigma}$ $=$ 289.76 pm/V, which is about
five times larger in magnitude than the clamped one. On the other
hand, the values of $r_{21}^{\sigma}$ $=$ 7.05 pm/V and $r_{31}^{\sigma}$
$=$ $-$12.66 pm/V are smaller than the clamped one because of the
negative piezoelectric contribution. The unclamped EO coefficient
of $r_{11}^{\sigma}$ $=$ 133.63 pm/V is also large in NbOI$_{2}$
monolayer, that is about four times stronger than that of the clamped
one. The remaining two unclamped EO coefficients in the monolayer
case, $r_{21}^{\sigma}$ $=$ 4.96 pm/V and $r_{31}^{\sigma}$ $=$
3.48 pm/V, are smaller in magnitude than their clamped values due
to the negative piezoelectric contribution too.

Let us also indicate that Equation (\ref{eq:elasto-optic coefficient})
can be used to obtain accurate elasto-optic coefficients from first-principles
calculations. For instance, a first-principles scheme predicted a
value of $p_{31}$ $=$ 0.17 for bulk LiNbO$_{3}$ \cite{Chen2015},
which agrees remarkably well with the experimental value of 0.18 \cite{Weber2002}.
The presently computed elasto-optic coefficients in NbOI$_{2}$ bulk
are $p_{11}$ $=$ 1.65, $p_{21}$ $=$ 0.61, and $p_{31}$ $=$ 0.39,
respectively. The predicted value of $p_{11}$ is therefore very large,
namely about 4 times larger in magnitude than that measured in tetragonal
BaTiO$_{3}$ with $p_{11}$ $=$ 0.425 \cite{Weber2002}. The magnitude
of the predicted $p_{31}$ is more than two times larger in magnitude
than that measured in bulk LiNbO$_{3}$ \cite{Weber2002}. The rescaled
elasto-optic coefficients in NbOI$_{2}$ monolayer are also large,
with $p_{11}$ $=$ 1.58, $p_{21}$ $=$ 0.46, and $p_{31}$ $=$
0.39, respectively, which further calls for the use of NbOI$_{2}$
in technologies taking advantage of elasto-optic conversions.

\begin{figure}
\includegraphics[width=8.5cm]{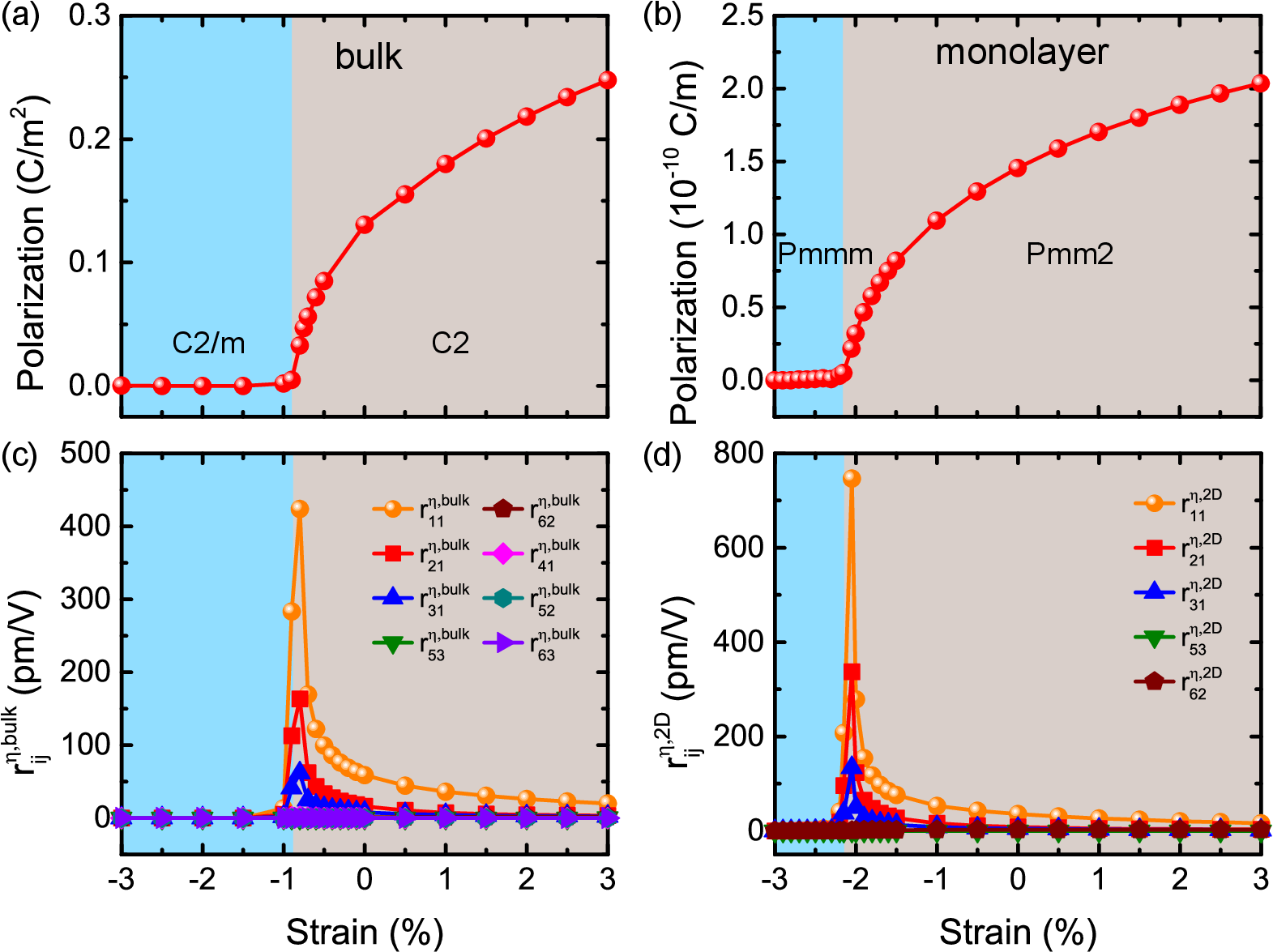}

\caption{The polarization $P_{x}$ as a function of strain in ferroelectric
NbOI$_{2}$ (a) bulk and (b) monolayer, respectively. The clamped
EO coefficients as a function of strain in NbOI$_{2}$ (c) bulk and
(d) monolayer, respectively. \label{fig:EO coefficients}}
\end{figure}

Let us now investigate the effect of strains on some properties in
these promising materials. Figure~\ref{fig:EO coefficients} shows
the polarization and clamped EO coefficients as a function of biaxial
epitaxial strain in NbOI$_{2}$ bulk and monolayer, respectively.
As mentioned above, a spontaneous polarization $P_{x}$ is along the
$x$ direction under stress-free conditions in both NbOI$_{2}$ bulk
and monolayer. It is computed from the Berry phase method \cite{King-Smith1993,Resta1994}.
Note that the calculated spontaneous polarization in NbOI$_{2}$ monolayer
is equal to 145.4 pC/m under stress-free conditions, which is in excellent
agreement with previous theoretical values of 142.5 pC/m \cite{Ye2021}
and 143 pC/m \cite{Wu2022}. The behaviors of the polarization and
EO coefficients allow the determination of two strain regions for
both bulk and monolayer. For strains ranging between $+$3\% and $-$0.8\%
in NbOI$_{2}$ bulk {[}Fig.~\ref{fig:EO coefficients}(a){]}, the
polarization gradually decreases from 0.25 to 0.03 C/m$^{2}$, with
the phase retaining its $C2$ space group. In contrast, the phase
for strains between $-$0.9\% and $-$3\% has the paraelectric $C2/m$
space group, with therefore no polarization and no finite EO coefficients.
For NbOI$_{2}$ monolayer, the polarization associated with the ferroelectric
phase (that has a $Pmm2$ space group) in the range of $+$3\% to
$-$2.05\% decreases from 2.0 $\times$ 10$^{-10}$ to 0.2 $\times$
10$^{-10}$ C/m {[}see Fig.~\ref{fig:EO coefficients}(b){]}. The
polarization and EO coefficients are null for strains between $-$2.1\%
to $-$3\%, since the resulting phase adopts the paraelectric $Pmmm$
space group.

Let us now pay closer attention to the EO response as a function of
strain in NbOI$_{2}$ bulk and monolayer, as shown in Figs.~\ref{fig:EO coefficients}(c)
and \ref{fig:EO coefficients}(d). At the boundary between ferroelectric
and paraelectric phases in NbOI$_{2}$ bulk, large values of the clamped
EO coefficients $r_{11}^{\textrm{\ensuremath{\mathrm{\eta,bulk}}}}$
and $r_{21}^{\textrm{\ensuremath{\mathrm{\eta,bulk}}}}$ are predicted
(more than 100 pm/V) due to the strain-driven occurrence of a phase
transition from $C2$-to-$C2/m$. For the clamped EO coefficients
in NbOI$_{2}$ monolayer, $r_{11}^{\textrm{\ensuremath{\mathrm{\eta,2D}}}}$
and $r_{21}^{\textrm{\ensuremath{\mathrm{\eta,2D}}}}$ also show very
large values near the boundary between $Pmm2$ and $Pmmm$ phases.
Strikingly, such large EO coefficients near this critical strain in
both bulk and monolayer are larger than the experimental values in
LiNbO$_{3}$ \cite{Weber2002,Turner1966} by one order of magnitude---strongly
suggesting to employ strained NbOI$_{2}$ materials for unprecedented
electro-optic devices performance. 
\begin{figure}
\includegraphics[width=8.5cm]{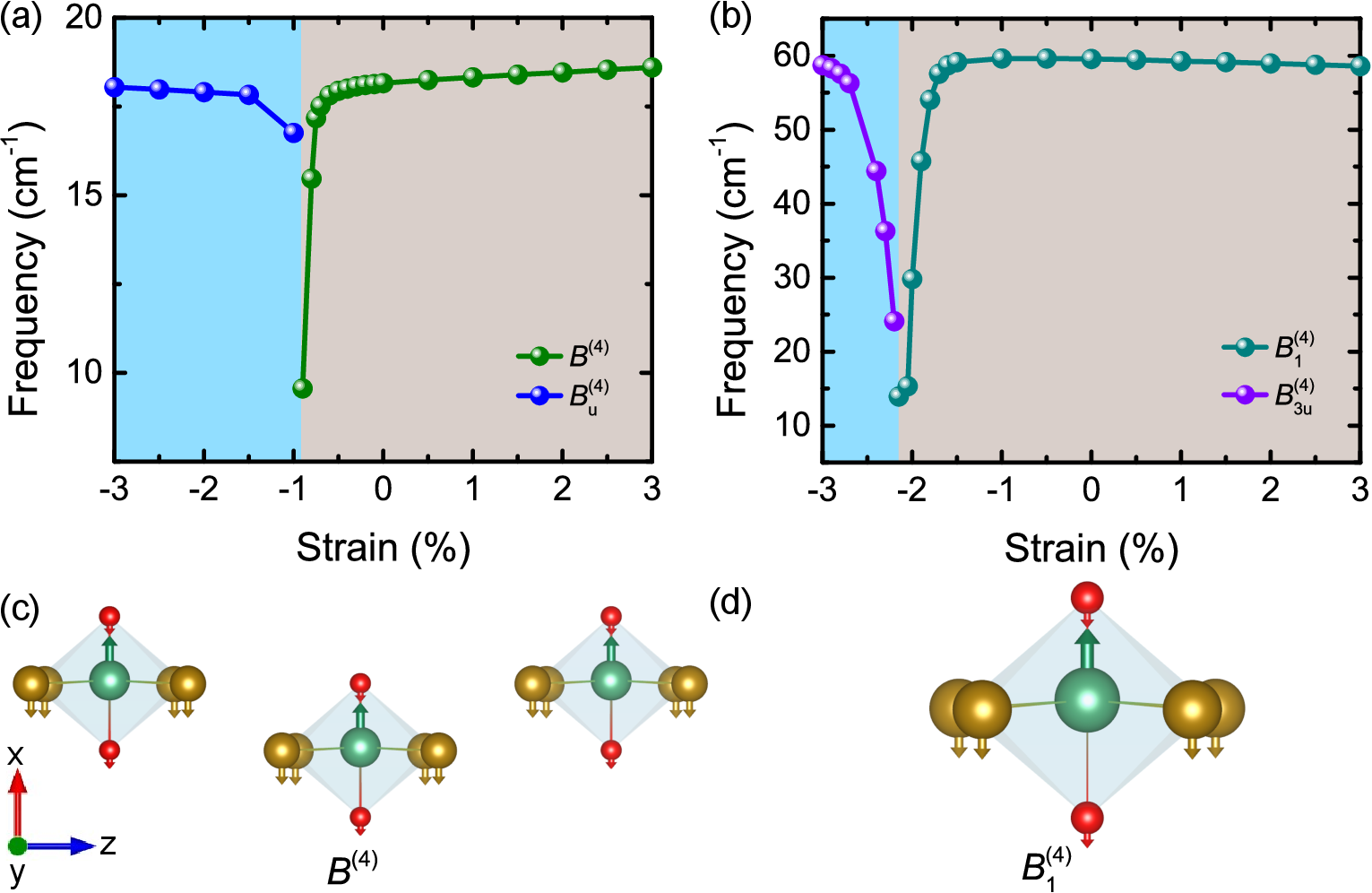}

\caption{The frequency of selected phonon modes as a function of strain in
ferroelectric (a) bulk and (b) monolayer NbOI$_{2}$, respectively.
Panels (c) and (d) show the atomic character of the eigenvector of
modes $B^{(4)}$ and $B_{1}^{(4)}$ in NbOI$_{2}$ bulk and monolayer,
respectively. \label{fig:frequency}}
\end{figure}

Four lowest phonon modes, expressed as $B_{u}^{(4)}$, $B^{(4)}$,
$B_{3u}^{(4)}$, and $B_{1}^{(4)}$ {[}see Figs.~\ref{fig:frequency}(a)
and \ref{fig:frequency}(b){]} are important in NbOI$_{2}$ bulk and
monolayer. As a matter of fact, for NbOI$_{2}$ bulk, Fig.~\ref{fig:frequency}(a)
reveals that the $B^{(4)}$ phonon mode becomes very soft around the
strain $\sim$ $-$0.8\%, which corresponds to the boundary between
$C2$ (ferroelectric) and $C2/m$ (paraelectric) phases. Having some
frequencies approaching zero is a guarantee to have large EO response,
as indicated by the second term of Eq.~(\ref{eq:EO_tensor_supercell}).
Indeed, we numerically found that the main contribution of phonon
modes for the largest $r_{11}^{\textrm{\ensuremath{\mathrm{\eta,bulk}}}}$
EO coefficient mostly arises from the polar mode $B^{(4)}$. Figure~\ref{fig:frequency}(c)
displays the atomic displacement of such mode at $-$0.8\% strain,
with the Nb ions displacing along the {[}100{]} direction while O
and I ions move along the opposite {[}$\bar{1}$00{]} direction in
NbOI$_{2}$ bulk.

For NbOI$_{2}$ monolayer, the continuous evolution from the ferroelectric
$Pmm2$ phase to the paraelectric $Pmmm$ state is driven by a softening
of the $B_{1}^{(4)}$ and $B_{3u}^{(4)}$ modes. At the phase boundary
of $-$2.05\% strain, the largest $r_{11}^{\textrm{\ensuremath{\mathrm{\eta,2D}}}}$
coefficient now mostly stems from the polar mode we denote as $B_{1}^{(4)}$,
which shows a similar atomic character of eigenvector {[}see Fig.~\ref{fig:frequency}(d){]}
as the mode $B^{(4)}$ of bulk case. One can thus safely conclude
that, in both strained NbOI$_{2}$ bulk and monolayer, the large linear
EO coefficients mainly originate from a strain-induced softening of
the lowest polar mode and a resulting ferroelectric-to-paraelectric
phase transition.

In summary, we investigated linear electro-optic and elasto-optic
effects in NbOI$_{2}$ bulk and monolayer from first-principles calculations.
We predict large clamped and unclamped EO and elasto-optic coefficients
in both stress-free bulk and monolayer. We also revealed the effect
of epitaxial strain on the EO response of NbOI$_{2}$ bulk and monolayer.
In particular, in both systems, a strain-induced ferroelectric-to-paraelectric
phase transition is discovered, being driven by a softening of some
lowest phonon modes and which results in very large linear EO responses.
The phase transitions and large EO responses are accompanied by change
in the atomic bond length and electronic band gap (see the Supplemental
Material (SM) \cite{SM}). The SM \cite{SM} further demonstrates
that other NbOX$_{2}$ materials, namely NbOBr$_{2}$ and NbOCl$_{2}$,
can also hold these spectacular effects. We compare the electro-optic
and elasto-optic coefficients in NbOI$_{2}$ bulk and monolayer with
other ferroelectric materials \cite{SM} and find that NbOI$_{2}$
has the best performance. Note that the phonon dispersions and stability
are also discussed in the SM \cite{SM}. We thus hope that the present
study will encourage the experimental investigation of electro-optic
and elasto-optic effects in bulk and 2D ferroelectric niobium oxide
dihalides. 
\begin{acknowledgments}
This work is supported by the National Natural Science Foundation
of China (Grant No.\ 12374092), Natural Science Basic Research Program
of Shaanxi (Program No.\ 2023-JC-YB-017), Shaanxi Fundamental Science
Research Project for Mathematics and Physics (Grant No.\ 22JSQ013),
``Young Talent Support Plan'' of Xi'an Jiaotong University (Grant
No.\ WL6J004), the Open Project of State Key Laboratory of Surface
Physics (Grant No.\ KF2023\_06), the Fundamental Research Funds for
the Central Universities, and the HPC Platform of Xi'an Jiaotong University.
C.P. acknowledges partial support through Agence Nationale de la Recherche
through Grant Agreement No. ANR-21-CE24-0032 (SUPERSPIN). C.P. and
L.B. thank the Defense Advanced Research Projects Agency Defense Sciences
Office (DARPA-DSO) Program: Accelerating discovery of Tunable Optical
Materials (ATOM) under Agreement No. HR00112390142 and the Award No.
FA9550-23-1-0500 from the U.S. Department of Defense under the DEPSCoR
program. L.B. also acknowledges the MonArk NSF Quantum Foundry supported
by the National Science Foundation Q-AMASE-i Program under NSF Award
No. DMR-1906383, the ARO Grant No. W911NF-21--1--0113, and the Vannevar
Bush Faculty Fellowship (VBFF) Grant No. N00014-20-1--2834 from the
Department of Defense. 

Z.Z. and X.D. contributed equally to this work.
\end{acknowledgments}

\end{document}